\documentclass[aps,pra,twocolumn,superscriptaddress,groupedaddress]{revtex4}

\usepackage{amssymb} \usepackage{color,graphicx} 
\usepackage{amsmath}
\usepackage{amsbsy} \usepackage{amsthm} \usepackage{bbm}
\usepackage{bm,bbm} \usepackage{float} \usepackage{braket}
\usepackage{placeins}
\usepackage[colorlinks=true,citecolor=blue,linkcolor=red,urlcolor=red]{hyperref}
\usepackage[sort&compress]{natbib}
\usepackage{comment}
\usepackage{cancel}
\usepackage{multirow}
\newcommand{\bq}{\begin{equation}} \newcommand{\eq}{\end{equation}}
\newcommand{\bqali}{\begin{equation}\begin{aligned}}
\newcommand{\eqali}{\end{aligned}\end{equation}}
\newcommand\D{\operatorname{d}\!}

\newcommand\rC{r_\text{\tiny C}}

\graphicspath{{./Figures/}}

\begin{document}

\begin{abstract}
{The experimental observation of a clear quantum signature of gravity is believed to be out of the grasp of current technology. However, several recent promising proposals to test the possible existence of non-classical features of gravity seem to be accessible by the state-of-art table-top experiments. Among them, some aim at measuring the gravitationally induced entanglement between two masses which would be a distinct non-classical signature of gravity. 
We explicitly study,  in two of these proposals, the effects of decoherence on the system's dynamics by monitoring the corresponding degree of entanglement. We identify the required experimental conditions necessary to perform successfully the experiments. In parallel, we account also for the possible effects of the Continuous Spontaneous Localization (CSL) model, which is the most known among the models of spontaneous wavefunction collapse. We find that any value of the parameters of the CSL model would completely hinder the generation of gravitationally induced entanglement.}
\end{abstract}

\title{Decoherence effects in non-classicality tests of gravity}
\author{Simone Rijavec}
\affiliation{Clarendon Laboratory, University of Oxford, Parks Road,Oxford OX1 3PU, United Kingdom}

\author{Matteo Carlesso}
\affiliation{Department of Physics, University of Trieste, Strada Costiera 11, 34151 Trieste, Italy}
\affiliation{Istituto Nazionale di Fisica Nucleare, Trieste Section, Via Valerio 2, 34127 Trieste, Italy}

\author{Angelo Bassi}
\affiliation{Department of Physics, University of Trieste, Strada Costiera 11, 34151 Trieste, Italy}
\affiliation{Istituto Nazionale di Fisica Nucleare, Trieste Section, Via Valerio 2, 34127 Trieste, Italy}

\author{Vlatko Vedral}
\affiliation{Clarendon Laboratory, University of Oxford, Parks Road,Oxford OX1 3PU, United Kingdom}
\affiliation{Centre for Quantum Technologies, National University of Singapore, 3 Science Drive 2, Singapore 117543}
\affiliation{Department of Physics, National University of Singapore, 2 Science Drive 3, Singapore 117542}

\author{Chiara Marletto}
\affiliation{Clarendon Laboratory, University of Oxford, Parks Road,Oxford OX1 3PU, United Kingdom}
\affiliation{Centre for Quantum Technologies, National University of Singapore, 3 Science Drive 2, Singapore 117543}
\affiliation{Department of Physics, National University of Singapore, 2 Science Drive 3, Singapore 117542}

\date{\today}

 \maketitle

\section{Introduction}

Testing the quantumness of gravity represents an outstanding challenge that has been approached from different perspectives \cite{kiefer2007}. Clearly the direct detection of the graviton - the quantum mediator of the gravitational interaction - would undeniably confirm the quantumness of gravity. However, the current state-of-art technology is not yet sufficiently advanced to allow for such a detection \cite{rothman2006,dyson2012}, and alternative paths need to be explored. Several low-energy table-top experiments were proposed over the years, which aim at testing effects resulting from either of a quantum \cite{pikovski2012,carney2019,carlesso2019,balushi2018,parikh2020} or a classical theory of gravity \cite{kafri2014}. Among such proposals, we focus on two of them, one by Bose \textit{et al.} \cite{bose2017} and Marletto-Vedral \cite{marletto2017c} (BM) and one by Krisnanda \textit{et al.}~\cite{krisnanda2020}, which seem to be within reach of the current technology.
They rely on the generation of a gravitationally induced entanglement between two masses, which would work as a witness of the non-classical nature of the gravitational mediator \cite{marletto2017c,bose2017}.
This new approach to witnessing non-classical features of gravity has been subject of intense study and debate {\cite{belenchia2018,altamirano2018,khosla2018,hall2018,anastopoulos2018,christodoulou2019,belenchia2019,marletto2017b,marletto2020,marshman2020,marletto2017a,bhole2020,chevalier2020,miki2021}}, and is at the core of several recent experimental proposals
 \cite{christodoulou2018,christodoulou2020,miao2020,howl2020,matsumura2020}.
Since the gravitational interaction is weak, one needs to employ large masses to achieve a measurable amount of entanglement in an experimentally reasonable time. However, larger masses are strongly affected by the environmental noises. Such effects, in particular decoherence, suppress quantum superpositions and thus hinder the entanglement generated by gravity. Therefore, it is crucial to consider explicitly environmental decoherence effects in the dynamics of the system. 
Moreover, since these setups are very sensitive to any source of decoherence, one should account also for non-standard decoherence sources as those described by models of spontaneous wavefunction collapse. These models represent  possible solutions to the quantum measurement problem. They  modify the standard evolution due to quantum mechanics by introducing stochastic and non-linear terms in the Schr\"odinger equation \cite{bassi2003,bassi2013}. While these models are still under active testing \cite{Adler:2019aa,Zheng:2020aa,vinante2020,Vinante:2020aa}, it is instructive to account for their possible effects in these setups.
Additional decoherence effects can rise from different processes related to gravity, as those proposed and analysed in \cite{terashima2004,terashima2005,esfahani2007,ahmadi2012,ahmadi2014,ralph2014,gooding2014,pikovski2015,gooding2015,carlesso_decoherence2016,pang2016,plato2016,dehdashti2017}. Our approach can easily be modified to include them, and the analysis is left for future research.

In this work, by building on the first analysis performed in \cite{nguyen2020,vandekamp2020} for the setup of Bose \textit{et al.}, we study the decoherence effects on the entanglement allegedly induced by non-classical gravity and identify the experimental conditions required to perform successfully the BM \cite{bose2017,marletto2017c} and Krisnanda \textit{et al.} \cite{krisnanda2020} proposals.
Moreover, we also quantify the effective decoherence due to the action 
of the Continuous Spontaneous Localization (CSL) model \cite{ghirardi1990,ghirardi1995}, which is the most studied among the models of spontaneous wavefunction collapse.

\section{WITNESSING NON-CLASSICAL FEATURES OF GRAVITY}\label{2}

The proposals in Refs.~\cite{bose2017,marletto2017c} and that in Ref.~\cite{krisnanda2020} to test non-classical features of gravity are based on the generation of gravitationally induced entanglement. Indeed, a classical mediator cannot induce entanglement between two systems that are not directly interacting \cite{horodecki2009}. Therefore, the observation of gravitationally induced entanglement between two massive systems implies that the gravitational mediator displays non-classical features. We briefly review the two proposals.
\subsection{BM proposal} 
\begin{figure}[t!]
	\includegraphics[width=0.4\linewidth]{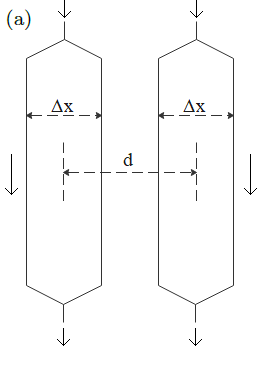}\hspace{0.05\linewidth}\includegraphics[width=0.5\linewidth]{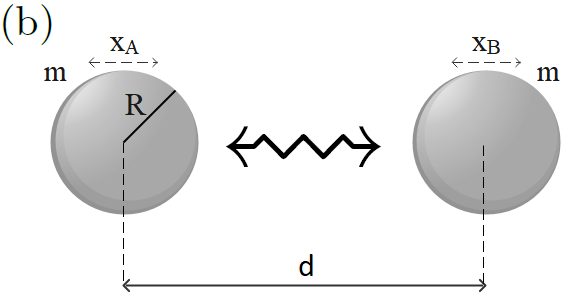}
	\caption{{Schematic representation of the two setups presented in Sec.~\ref{2}.
	(a) Setup proposed 
	by Bose \textit{et al.} \cite{bose2017} and Marletto-Vedral \cite{marletto2017c}. Two diamond particles with an embedded spin are initially prepared in the state $(\ket{\uparrow}+\ket{\downarrow})/\sqrt{2}$ and subsequently sent through two adjacent Stern-Gerlach interferometers. 
	 The particles are initially at a distance $d$ and the size of the superpositions in the interferometer is $\Delta x$.}
	(b) Setup proposed by Krisnanda \textit{et al.} \cite{krisnanda2020}. Two osmium spheres of radius $R$ and mass $m$ are initially trapped in two harmonic potentials with frequency $\omega_0$ at a distance $d$. The spheres are initially cooled down close to the ground state of the harmonic potentials and the traps are subsequently removed, thus letting the masses freely interact. Here,  $\hat x_\text{\tiny A}$ and $\hat x_\text{\tiny B}$ denote the displacement of the centres of mass of the two spheres from their initial equilibrium position.}
	\label{setup_bose}
\end{figure}
The first setup proposed to test gravitationally-induced entanglement is that of Bose \textit{et al.} \cite{bose2017} and Marletto-Vedral \cite{marletto2017c} (BM). The setup is based on Stern-Gerlach interferometry \cite{machluf2013,wan2016,amit2019}, however alternative experimental schemes, as for example that exploiting magnetic levitation  \cite{pino2018}, can be also considered. Two diamond particles of mass $m$ and radius $R$ embedded with a single spin $1/2$ are initially prepared in the  superposition state $(\ket{\uparrow}+\ket{\downarrow})/\sqrt{2} $ and sent simultaneously through two Stern-Gerlach interferometers as depicted in Fig.~\ref{setup_bose}(a). After passing through an inhomogenous magnetic field, the spin superposition induces a position superposition for each of the particles, thus entangling the spin and the position. The composite state of the system thus becomes  
\bq\label{initialstate}
\ket{\psi}_1\otimes\ket{\psi}_2\, \quad\text{where}\quad\ket{\psi}_i=\frac{\ket{\uparrow,\text{ L}}_i+\ket{\downarrow,\text R}_i}{\sqrt{2}}
\eq
is the single particle state, and $\ket{L}$ and $\ket{R}$ are the localized states of the particles in the left and right branch of the interferometer, respectively. This assumption about the states of the particles has been shown to be a valid approximation for this setup \cite{chevalier2020}. 

Now gravity comes in play, mediating the interaction between the two masses. If gravity is to obey the quantum superposition principle, it induces a different phase for each of the parts of the superposition, which depends on the distance from the other branches of the superposition \cite{bose2017}. 
This interaction leads to an entangled state of the two masses \cite{bose2017,marletto2017c}.

An estimate of the gravity-induced phases can be obtained by assuming that
 the dominant effect can be computed via Newtonian interaction \cite{marletto2017c,bose2017,belenchia2018,christodoulou2019}; this approximation is valid also in the linearised quantum gravity model \cite{marletto2018,kiefer2007}.
In particular,  
with reference to the setup represented in Fig.~\ref{setup_bose}(a), we assume that the gravitational interaction is sufficiently weak to not modify the relative distances between the branches but at the same time strong enough to modify the corresponding relative phases. 
Moreover, we assume that all other interactions among the two particles are negligible and we account for
the action of gravity only in the parallel branches of the interferometers.
In such a way, the problem is significantly simplified. The corresponding driving Hamiltonian, in the  $\{\ket{i}_1\otimes\ket{j}_2\}$ with $i,j=\text{L}, \text{R}$ basis representation, reads
\begin{equation}
H_\text{\tiny BM-g}=\begin{pmatrix}
U_{0} & 0 & 0 & 0 \\
0 & U_{-} & 0 & 0 \\
0 & 0 & U_{+} & 0 \\
0 & 0 & 0 & U_{0} \\
\end{pmatrix} ,
\label{H_bose}
\end{equation}
where we defined 
\begin{align}
U_0=G\frac{m^2}{d},\quad U_{\pm}=G\frac{m^2}{d\mp \Delta {x}},
\end{align}
with $d$ denoting the distance between the center of mass of the two particles and $\Delta x$ the superposition distance.
By assuming that the initial state is that in Eq.~\eqref{initialstate}, the corresponding density matrix at time $t$ reads:
\begin{equation}
\rho=\frac{1}{4}\begin{pmatrix}
1 & e^{-i\Delta_{0-}t} & e^{-i\Delta_{0+}t} & 1 \\
e^{i\Delta_{0-}t} & 1 & e^{-i\Delta_{-+}t} & e^{i\Delta_{0-}t} \\
e^{i\Delta_{0+}t} & e^{i\Delta_{-+}t} & 1 & e^{i\Delta_{0+}t} \\
1 & e^{-i\Delta_{0-}t} & e^{-i\Delta_{0+}t} & 1 \\
\end{pmatrix} ,
\end{equation}
with
\begin{equation}
\Delta_{ij}=\frac{U_i-U_j}{\hbar}\,,\text{ with } i,j\in \{0,+,-\} .
\end{equation}
The entanglement of the density matrix $\rho$ is verified by applying the Peres-Horodecki criterion \cite{peres1996,horodecki1997}. In particular,
 we calculate the eigenvalues $\tilde{\lambda}_i$ of the partially transposed density matrix, which we explicitly derive in Appendix \ref{appA}. They read:
\begin{equation}
\tilde{\lambda}_{1}^{\pm}=\frac{1}{2}  \left(1\pm \left|\text{cos}\,\frac{t}{\tau_{G}}\right|\right),  \hspace{0.5cm}
\tilde{\lambda}_{2}^{\pm}=\pm \frac{1}{2} \left|\text{sin}\,\frac{t}{\tau_{G}}\right|,
\label{eigen_bose_free}
\end{equation}
where we defined the characteristic time
\begin{equation}
\tau_{G}
=	\frac{\hbar d \left[\left(\frac{d}{\Delta x}\right)^2-1\right]}{ G m^2}.
\label{T_bose}
\end{equation}
According to the  Peres-Horodecki criterion, the system is in an entangled state if at least one among the eigenvalues $\tilde{\lambda}_i$ is negative. The condition is both necessary and sufficient for $2\times2$ dimensional systems as in our case. 
We notice that $\tilde{\lambda}_2^-$ is always negative $\forall t\neq k\pi \tau_{G}$ with $k \in \mathbb{Z}$.
We quantify the corresponding entanglement by exploiting the logarithmic negativity $E=\log_2||\tilde{\rho}||_1$, where $||\cdot||_1$ is the trace norm \cite{horodecki2009}. This gives
\begin{equation}
E_\text{\tiny BM-free}=\log_2 \left(1+ \left|\text{sin}\,\frac{t}{\tau_{G}}\right|\right).
\end{equation}
The entanglement of the system has a periodic dynamics and reaches its first maximum value at time $t_\text{\tiny MAX}=\tfrac{\pi}{2} \tau_{G}$.
By inserting the values of the parameters chosen by Bose \textit{et al.} \cite{bose2017}, which are reported in Tab.~\ref{par}, we have $t_\text{\tiny MAX}\simeq 25\,$s.
We remark that the value of $d$ has been chosen by Bose \textit{et al.} such that the Casimir-Polder interactions between the two spheres are 10 times smaller than the gravitational interaction. In this way, it is easier to discriminate the contribution of gravity to entanglement from that of these short-range forces. A slightly different setup with mitigated Casimir-Polder interactions has been recently proposed in \cite{nguyen2020} and thoroughly analysed in \cite{vandekamp2020}. This setup proposes to insert a conducting plate between the two
spheres and provides an improvement of one/two orders of magnitude on the mass and superposition size with respect to \cite{bose2017}. The corresponding analysis of decoherence acting on the experimental apparatus and that due to external acceleration noises were performed in \cite{toros2020} and \cite{grossardt2020} respectively, while the decoherence effects on the system were not quantified. In an attempt to simplify the treatment of the decoherence effects, here we will not consider this modified version of the setup, although our analysis can be applied also to this scenario.

\begin{table}[t]
	\caption{Numerical value of the parameters proposed in the setups
	by Bose \textit{et al.} \cite{bose2017}, Marletto-Vedral \cite{marletto2017c} and by Krisnanda \textit{et al.} \cite{krisnanda2020}. The masses $m$ are assumed to be spheres of radius $R$ and are separated by a distance $d$. For the BM setup one has a superposition distance of $\Delta x$, while in that of Krisnanda \textit{et al.} the masses are initially confined in an harmonic trap of frequency $\omega_0$}
	\label{par}
	\centering
{\small
\begin{tabular}{c|ccccc}
\hline\hline
Proposal&$m$ [kg]&$R$ [m]&$d$ [m]&$\Delta x$ [m]&$\omega_0$ [Hz]\\
\hline
Bose &$10^{-14}$&$10^{-6}$&$4.5\times10^{-4}$&$2.5\times10^{-4}$&\cancel{\phantom{$1$}}\\
Marletto&$10^{-12}$& \cancel{\phantom{$1$}} &$10^{-6}$& \cancel{\phantom{$1$}} &\cancel{\phantom{$1$}}\\
Krisnanda &$10^{-7}$&$10^{-4}$&$3\times10^{-4}$&\cancel{\phantom{$1$}}&$10^5$\\
\hline\hline
\end{tabular}
}
\end{table}

\subsection{Krisnanda \textit{et al.} proposal}
The second setup we consider was recently proposed by Krisnanda \textit{et al.} in \cite{krisnanda2020}.
The setup is schematically depicted in Fig.~\ref{setup_bose}(b) and consists of two osmium spheres of mass $m$ and radius $R$ that are initially trapped in harmonic potentials with frequency $\omega_0$ separated by a distance $d$. The masses are assumed to be cooled down close to the ground state of their harmonic potentials, which is achievable with current technology \cite{chan2011,teufel2011}. The particles are subsequently released from the traps and are let free to interact gravitationally and entangle. Again, the Newtonian potential can be exploited to effectively describe such an interaction.
The quantification of the degree of entanglement can be provided through well established continuous variables techniques applied to bipartite Gaussian states \cite{serafini2017}. In particular, in such a case, the separability of the state can be simply verified through its $4\times4$ covariance matrix in position and momentum. 

In the proposal of Krisnanda \textit{et al.}, one assumes that the two masses are initially prepared in Gaussian thermal states of the harmonic potentials. Moreover, one can approximate the  Newtonian potential to second order in the quantum fluctuations $(\hat x_\text{\tiny A}-\hat x_\text{\tiny B})$ of the relative distance $d$ \cite{krisnanda2020}:
\begin{equation}
\hat{H}_\text{\tiny K-g}=-\frac{Gm^2}{d}\left(1+\frac{\left( \hat{x}_\text{\tiny A}-\hat{x}_\text{\tiny B} \right)}{d}+\frac{\left( \hat{x}_\text{\tiny A}-\hat{x}_\text{\tiny B} \right)^2}{d^2}\right),
\label{HG}
\end{equation}
 where the last term entangles the two masses \cite{krisnanda2020,qvarfort2020}. Under this approximation, which is valid for $|\hat{x}_A-\hat{x}_B|\ll d$ according to the parameters reported in Tab.~\ref{par} and for a duration of the experiment of around $1\,$s \cite{krisnanda2020}, the dynamics is  Gaussian.
Krisnanda \textit{et al.}~derived the analytic expression for the logarithmic negativity of the system in the limit ${2Gm}\ll{\omega_0^2 d^3}$  and $\sqrt{{2Gm}}\omega t \ll {\omega_0 d^{3/2}}$, which reads \cite{krisnanda2020}:
\begin{equation}
E_\text{\tiny K-free}=-\text{log}_2\left( \sqrt{1+2\sigma(t)-2\sqrt{\sigma(t)^2+\sigma(t)}} \right),
\end{equation}
where $\sigma(t)=4G^2m^2\omega_0^2t^6/9d^6$.
Using the parameters reported in Tab.~\ref{par}, after 1\,s, the logarithmic  negativity is $10^{-2}$, a value which can be detected with current technology \cite{palomaki2013}.

\section{DECOHERENCE EFFECTS}\label{3}

To measure the entanglement created by the quantum gravitational interaction one must ensure that environmental decoherence is sufficiently weak to not spoil such an effect. The main sources of decoherence are the scattering of residual gas and the scattering, emission and  absorption of thermal photons.
Here, we compute explicitly these effects  on the dynamics of the system in the two setups above described, and  their influence on the gravitationally-induced entanglement.

\subsection{BM proposal}
The effect of decoherence on the BM setup 
can be conveniently described in terms of a master equation, which in the position representation reads \cite{schlosshauer2007,romero-isart2011}:
\begin{equation}
\frac{\D\rho(\bm{x},\bm{x}',t)}{\D t}= -\frac{i}{\hbar} \bra{\bm{x}}\left[\hat{H},\hat{\rho}(t)\right]\ket{\bm{x}'} -\Gamma(|\bm{x}-\bm{x}'|) {\rho}(\bm{x},\bm{x}',t) ,
\label{master_deco}
\end{equation}
where $\hat H$ is the free Hamiltonian,
${\rho}(\bm{x},\bm{x}',t)= \braket{\bm{x}|\hat{\rho}(t)|\bm{x}'}$ and we consider the common ansatz \cite{romero-isart2011}
\begin{equation}
\Gamma(\Delta x)=\Gamma_0\left(1 - \text{exp}\left[-\frac{\Delta x^2}{4a^2}\right]\right).
\label{gamma_deco}
\end{equation}
Such a master equation  leads to an exponential suppression in time of the off-diagonal terms of the density matrix in the position representation.
According to Eq.~\eqref{gamma_deco}, decoherence is characterized by the  localization strength
$\Gamma_0$  and localization distance $a$, whose 
explicit forms are reported in Tab.~\ref{decoherence_parameters}. 
\begin{table}[b]
\caption{{Values of $a$ and $\Gamma_0$ entering in Eq.~\eqref{gamma_deco} which quantify the decoherence effects induced by collisions with air molecules (Air) and scattering (Sc), absorption (Ab) and emission (Em) of thermal photons on a sphere of radius $R$, dielectric constant $\epsilon$ and bulk temperature $\mathsf T_i$ \cite{schlosshauer2007,romero-isart2011}. 
Here, $m_\text{\tiny air}$ denotes the mass of the molecules of the residual air, $\mathsf T$ and $\mathsf P$ are respectively the temperature and the pressure at which the experiment is performed, and $\zeta(n)$ is the Riemann zeta function. To quantify the effects, one has $\epsilon=5.7+i\times10^{-4}$ \cite{derkach2005} for diamond used in the BM proposal, while $\epsilon=0.6+i\times 2.5$ \cite{Windt:1988aa} for osmium used in the Krisnanda proposal. Moreover, for the sake of simplicity, we take $\mathsf T_i=\mathsf T$ and $m_\text{\tiny air}\sim 6.6\times 10^{-27}\,$kg corresponding to a atom of helium.}}
\label{decoherence_parameters}
	\centering
	\begin{tabular}{c|cc}
		\hline
			\hline
		Source & $a^{i}$ & $\Gamma_0^i$ \\ 
		\hline
		{Air} & $\frac{\pi \hbar}{\sqrt{2\pi m_\text{\tiny air} k_\text{\tiny B} \mathsf T}}$ & $\frac{16\pi\sqrt{2\pi}}{3}\frac{\mathsf P R^2}{\sqrt{m_\text{\tiny air} k_\text{\tiny B} \mathsf T }}$ \\ 
		{Sc} & $\frac{\pi^{2/3}\hbar c}{2k_\text{\tiny B} \mathsf T}$  & $8!\frac{8\pi^{1/3}}{9} R^6c  \left(\frac{k_\text{\tiny B} \mathsf T}{\hbar c}\right)^7 \zeta(9) \text{Re}\left[\frac{\epsilon-1}{\epsilon+2}\right]^2$ \\
		{Ab} & $\frac{\pi^{2/3}\hbar c}{2k_\text{\tiny B} \mathsf T}$  & $\frac{16\pi^{19/3}}{189} R^3 c  \left(\frac{k_\text{\tiny B} \mathsf T}{\hbar c}\right)^4 \text{Im}\frac{\epsilon-1}{\epsilon+2}$  \\
		{Em} & $\frac{\pi^{2/3}\hbar c}{2k_\text{\tiny B} \mathsf T_i}$  & $\frac{16\pi^{19/3}}{189} R^3 c  \left(\frac{k_\text{\tiny B} \mathsf T_i}{\hbar c}\right)^4 \text{Im}\frac{\epsilon-1}{\epsilon+2}$  \\ \hline	\hline
	\end{tabular}
\end{table}
%
In the proposals under consideration, there are only four possible position configurations, thus the problem is strongly simplified to a discrete description. 
Moreover, we assume that the decoherence acts independently on the two masses.

\begin{figure}[t]
	\centering
	\includegraphics[scale=0.7]{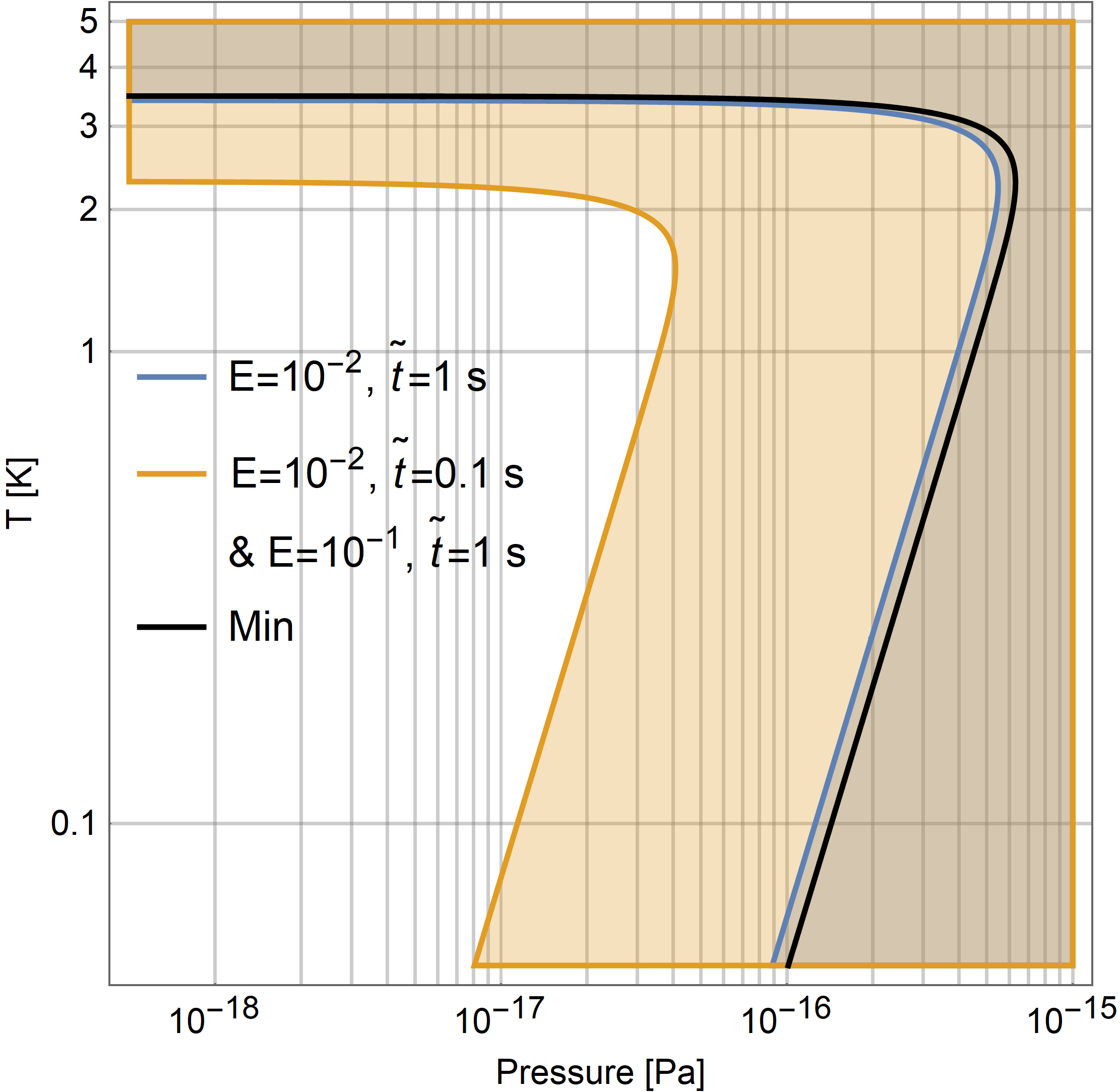}
	\caption{Environmental conditions required to observe different amounts of entanglement in the BM setup. Each line represents the necessary environmental conditions required to observe a specific value of the logarithmic negativity $E$ as reported in Eq.~\eqref{neg_bose} at time $\tilde{t}$. The respective shaded areas correspond to values of pressure and temperature that prevent  observing the targeted value of $E$ within time $\tilde{t}$. The black line identified with \textit{Min} denotes the minimum experimental conditions required to generate entanglement. We note that $E=10^{-2}$ is compatible with recent experimental observations \cite{palomaki2013}.  The residual gas is assumed to be helium.}
	\label{pT_bose}
\end{figure}

 By starting from the common initial state defined in Eq.~\eqref{initialstate},
the density matrix  $\rho$  of the system at time $t$ reads:
\begin{equation}
\rho=\frac{1}{4}\left(\begin{smallmatrix}
1 & e^{-i\Delta_{0-}t-\Gamma t} & e^{-i\Delta_{0+}t- \Gamma t} & e^{-2\Gamma t} \\
e^{i\Delta_{0-}t-\Gamma t} & 1 & e^{-i\Delta_{-+}t-2\Gamma t} & e^{i\Delta_{0-}t-\Gamma t} \\
e^{i\Delta_{0+}t-\Gamma t} & e^{i\Delta_{-+}t-2\Gamma t} & 1 & e^{i\Delta_{0+}t-\Gamma t} \\
e^{-2\Gamma t} & e^{-i\Delta_{0-}t-\Gamma t} & e^{-i\Delta_{0+}t-\Gamma t} & 1 \\
\end{smallmatrix} \right),
\end{equation}
where $\Gamma=\Gamma^\text{\tiny air}(\Delta x)+\Gamma^\text{\tiny ph,sc}(\Delta x)+\Gamma^\text{\tiny ph,abs}(\Delta x)+\Gamma^\text{\tiny ph,em}(\Delta x)$ is the sum of the effects due to the collisions of the residual air molecules and the scattering and absorption and emission of thermal photons. The entanglement is then quantified as in the free case via the Peres-Horodecki criterion \cite{peres1996,horodecki1997}. While the entire derivation is reported in Appendix \ref{appA}, here we show the obtained  eigenvalues $\tilde{\lambda}_i$ of the partially transposed density matrix: 
\begin{equation}
\begin{split}
&\tilde{\lambda}_{1}^{\pm}=\frac{e^{-\Gamma t}}{2} \left(\text{cosh}\,\Gamma t\pm \left|\text{cos}\,\frac{t}{\tau_{G}}\right|\right),  \\
&\tilde{\lambda}_{2}^{\pm}=\frac{e^{-\Gamma t}}{2}\left(\text{sinh}\,\Gamma t\pm \left|\text{sin}\,\frac{t}{\tau_{G}}\right|\right),
\end{split}
\label{eigen_bose_deco}
\end{equation}
where $\tau_{G}$ is defined in Eq. \eqref{T_bose}. We remind that entanglement can be achieved only for negative values of one of the above eigenvalues. Since $\cosh \Gamma t \geq 1$ and $\sinh \Gamma t \geq 0$, negative values can be achieved only by $\tilde{\lambda}_2^-$. This happens when 
\begin{equation}
	\Gamma<\frac{1}{\tau_{G}}=\frac{ G m^2}{\hbar d (\left(\frac{d}{\Delta x}\right)^2-1)},
	\label{env_bose}
\end{equation}
or, equivalently, when
\begin{equation}
\tau_C >  \tau_{G},
\label{tau_bose}
\end{equation}
where $\tau_C= 1/\Gamma$ is the coherence time of the system \cite{schlosshauer2007}.
The corresponding logarithmic negativity reads:
\begin{equation}
E_\text{\tiny BM-dec}=\max\left\{0,\text{log}_2 \left [e^{-\Gamma t} \left(\text{cosh}\,\Gamma t+ \left|\text{sin}\,\frac{t}{\tau_{G}}\right|\right)\right]\right\}.
\label{neg_bose}
\end{equation}
\begin{figure}[t]
	\centering	
	\includegraphics[scale=0.33]{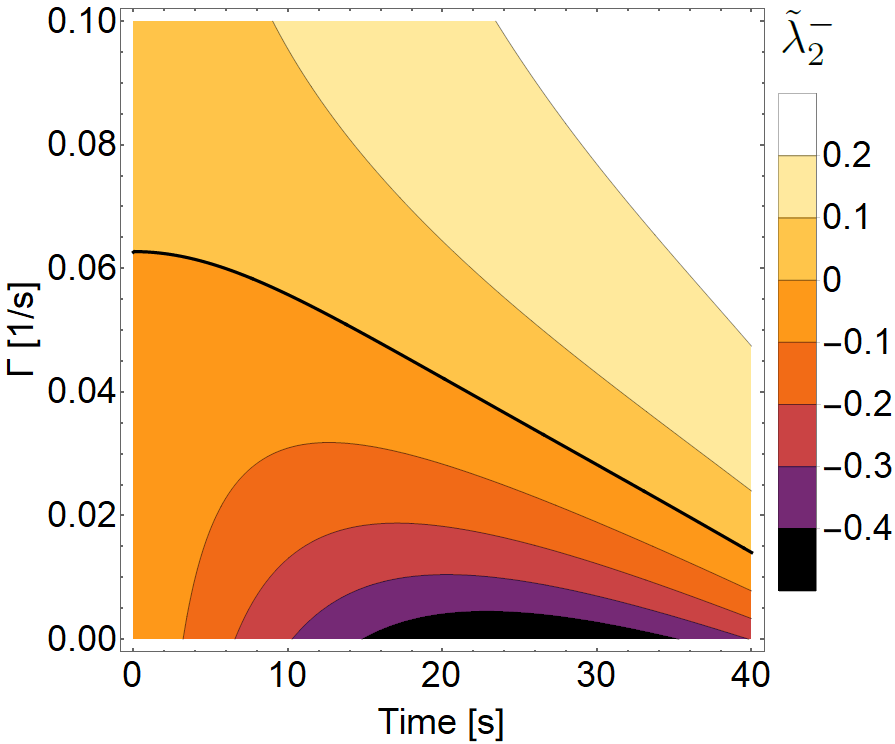}	
	\includegraphics[scale=0.5]{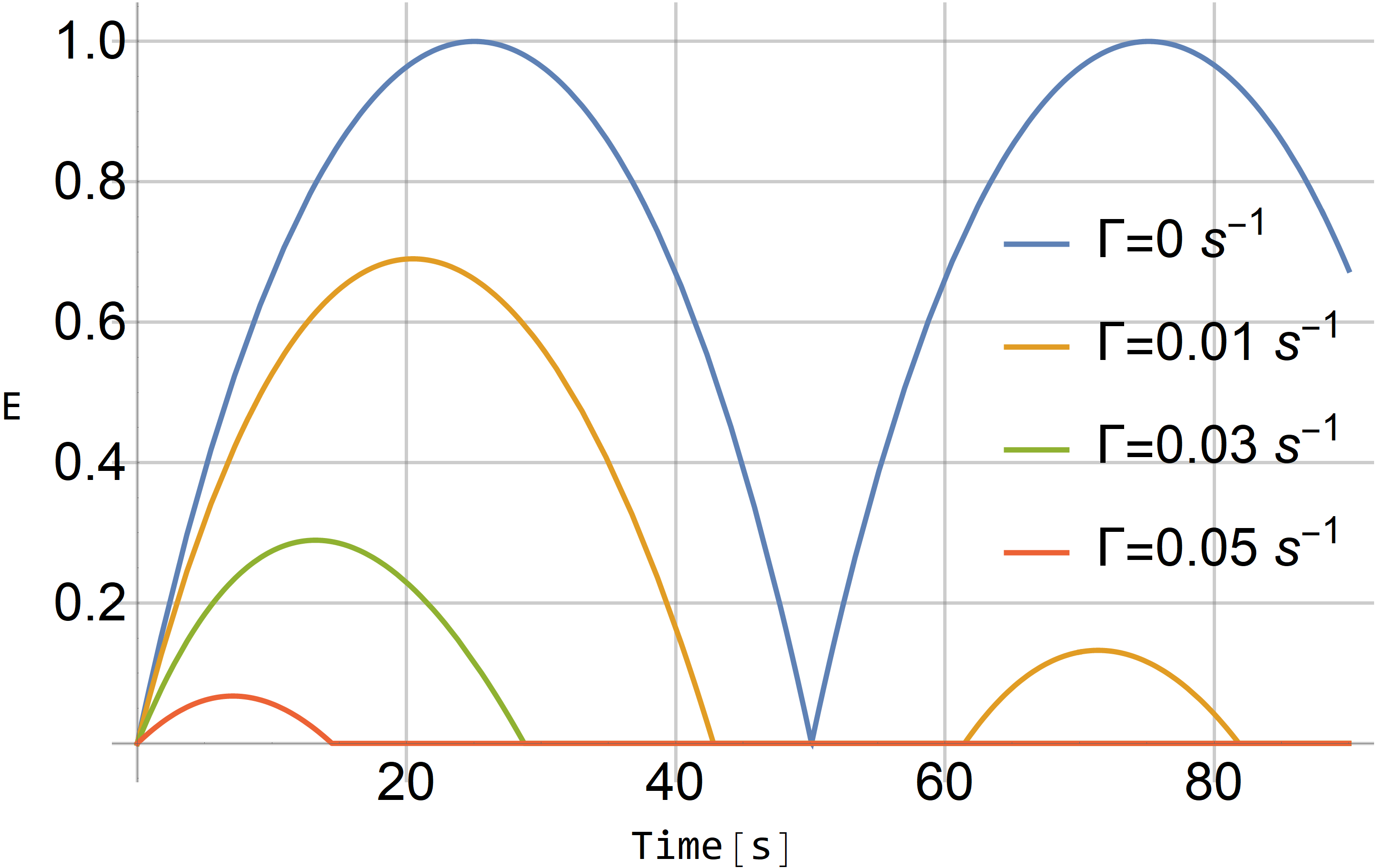}
	\caption{Entanglement indicators for the BM proposal.
		(Top panel)  The eigenvalue $\tilde{\lambda}_2^{-}$ provided by Eq. \eqref{eigen_bose_deco} as function of time and of the localization strength $\Gamma$. The region $\tilde{\lambda}_2^{-}<0$, which is below the black curve, corresponds to entangled states of the system.
		(Bottom panel) Evolution of the logarithmic negativity of the system $E$ given in Eq. \eqref{neg_bose} for different values of $\Gamma$.}
	\label{ent_bose}
\end{figure}
In the upper panel of Fig.~\ref{ent_bose}, we report  $\tilde{\lambda}_2^{-}$ for different values of time and of $\Gamma$. In the bottom panel, we report the logarithmic negativity of the system for different values of $\Gamma$.
In Fig.~\ref{pT_bose}, we have plotted the environmental conditions of temperature and pressure required to reach different amounts of entanglement with the parameters of Tab.~\ref{par}.  Given these parameters, we observe in Fig.~\ref{ent_bose} that for $\Gamma$ greater than $1/\tau_{G}\simeq 0.06\,$s$^{-1}$ the system is separable at all times.
 This corresponds to pressures and temperatures highlighted by the blue line named ``Min'' in Fig.~\ref{pT_bose}.
Moreover, we notice that one needs temperatures lower than $4\,$K and residual air pressures lower than $10^{-16}\,$Pa to detect entanglement in the setup. With these pressures and temperatures, considerable amounts of entanglement are reached in about $1\,$s. 

\subsection{Krisnanda \textit{et al.} proposal}

In the  setup of Krisnanda \textit{et al.}, the effect of environmental decoherence on the system can be described by the Heisenberg-Langevin equations for the position $\hat{x}_j$ and momentum $\hat{p}_j$ operators of the two masses. 
In one dimension, these equations read \cite{giovannetti2001}: 
\begin{equation}
\begin{aligned}
\frac{\D\hat{x}_j(t)}{\D t}& =  \frac{i}{\hbar}\left[\hat{H},\hat{x}_j(t)\right] , \\
\frac{\D\hat{p}_j(t)}{\D t}& =  \frac{i}{\hbar}\left[\hat{H},\hat{p}_j(t)\right] -\gamma_{\varepsilon} \hat{p}_j(t) +{\hat{\xi}_j(t)}  . \label{lang} 
\end{aligned}
\end{equation}
where $\hat H$ is the free Hamiltonian of the system, $\gamma_{\varepsilon}$ quantifies the dissipation of the environment and it is related to 
$\Gamma_0$ through \cite{breuer2002}
\begin{equation}
\gamma_{\varepsilon}=\frac{\Lambda \hbar^2}{m k_\text{\tiny B} \mathsf T},
\label{gamma_lambda}
\end{equation}
with $\Lambda= {\Gamma_0}/{4a^2}$, 
 $m$ denoting the mass of the particle and $ \mathsf T$ the temperature of the environment. 
$\hat{\xi}_j(t)$ is an environmental noise operator,
which can be described in terms of its mean
$\braket{\hat{\xi}_j(t)}_{\varepsilon}=0$, where $\braket{\cdot}_{\varepsilon}$ denotes the average over the environmental degrees of freedom, and the two-time correlation function. In particular, the latter can be strongly simplified in the Markovian regime which can be achieved in the limit of high temperatures of the environment: $k_\text{\tiny B} \mathsf T\gg\hbar \omega$. In such a case, it reads
 \cite{giovannetti2001,breuer2002}:
\begin{equation}
\frac{\braket{{\hat{\xi}_j}(t){\hat{\xi}_j}(t')+{\hat{\xi}_j}(t'){\hat{\xi}_j}(t)}_{\varepsilon}}{2} =2\hbar^2 \Lambda\delta (t-t'),
\label{corr_env}
\end{equation}
where $\delta(\cdot)$ is the Dirac delta.  Using the parameters considered by Krisnanda \textit{et al.}, which are reported in Tab. \ref{par}, we see that the high-temperature limit holds for $\mathsf T\gg 10^{-6}\,$K. 

We now apply Eq.~\eqref{lang} to the setup under consideration.
By denoting with $\omega_0$
the frequency of the harmonic potentials where the two particles are initially trapped, the Langevin equations for 
the adimensional quadratures $\hat{X}_j  = \sqrt{m\omega_0/\hbar}\,\hat{x}_j$ and $\hat{P}_j=\hat{p}_j/ \sqrt{(\hbar m \omega_0)}$, read  
\begin{align}
\frac{d\hat{X}_j}{dt}& =  \omega_0 \hat{P}_j \quad \text{with} \quad j=A,B, \nonumber\\
\frac{d\hat{P}_A}{dt}& =  \omega_0\eta\left( \hat{X}_A- \hat{X}_B\right) -\gamma_{\varepsilon}\hat{P}_A +\hat{\xi}^{\prime}_A +k , \label{lang_env} \\
\frac{d\hat{P}_B}{dt}& =  \omega_0\eta\left( \hat{X}_B- \hat{X}_A\right)-\gamma_{\varepsilon}\hat{P}_B +\hat{\xi}^{\prime}_B-k  , \nonumber
\end{align}
where we defined $\gamma_{\varepsilon}=\gamma^\text{\tiny air}+\gamma^\text{\tiny ph,sc}+\gamma^\text{\tiny ph,abs}+\gamma^\text{\tiny ph,em}$ and $\hat{\xi}^{\prime}_j=\left(\hat{\xi}_j^\text{\tiny air}+\hat{\xi}_j^\text{\tiny ph,sc}+\hat{\xi}_j^\text{\tiny ph,abs}+\hat{\xi}_j^\text{\tiny ph,em}\right)/\sqrt{\hbar m \omega_0}$, and  introduced the parameters
\begin{equation}
\eta= \frac{2Gm}{\omega_0^2 d^3},\quad\text{and}\quad k= \frac{Gm^2}{\sqrt{\hbar m \omega_0 d^4}}
\label{eta}
\end{equation}
which characterize the strength of the gravitational attraction \cite{krisnanda2020}.
We notice that, since the separation between the two spheres is much greater than the localization distances of the various processes, the noises acting on the two spheres can be safely considered as independent. Therefore, the two-time correlation function becomes
\begin{equation}
\frac{\braket{\hat{\xi}^{\prime}_{i}(t)\hat{\xi}^{\prime}_{j}(t')+\hat{\xi}^{\prime}_{j}(t')\hat{\xi}^{\prime}_{i}(t)}}{2} =\frac{2\hbar\Lambda }{m\omega_0}\delta (t-t') \delta_{ij} ,
\label{corr_TOT}
\end{equation}
where $\Lambda=\Lambda^\text{\tiny air}+\Lambda^\text{\tiny ph,sc}+\Lambda^\text{\tiny ph,abs}+\Lambda^\text{\tiny ph,em}$.

The procedure that we use to calculate the logarithmic negativity of the system in this setup is reported thoroughly in Appendinx \ref{appB} and is here briefly outlined.
First of all, we use the Langevin equations \eqref{lang_env} to compute the covariance matrix $\bm{\sigma}$ for the quadratures $\hat X_j$ and $\hat P_j$. Using this matrix we can easily find the covariance matrix $\tilde{\bm{\sigma}}$ of the partially transposed state of the system $\tilde{\rho}$. The entanglement of the system can be calculated by using the so-called symplectic eigenvalues $\tilde{\nu}_{i}$ of $\tilde{\bm{\sigma}}$. For bipartite Gaussian states, the minimum $\tilde{\nu}_\text{\tiny min}$ of these eigenvalues is smaller than $1/2$ if and only if the system is in an entangled state. Finally, the logarithmic negativity of the system can be expressed as \cite{serafini2017}
\bq
E_\text{\tiny K-dec}=-{\log}_2(2\tilde{\nu}_\text{\tiny min}).
\label{logneg}
\eq
The explicit expression for $\tilde{\nu}_{min}$ is cumbersome and we report it in Appendix \ref{appB}. Here, we report its behaviour for small times:
\begin{equation}
\tilde{\nu}_\text{\tiny min}=\frac{1}{2}+\frac{1}{2}\left(\frac{2\hbar\Lambda }{m\omega_0}-\eta\omega_0-\gamma_{\varepsilon}\right)t+\mathcal O(t^2).
\label{nu0}
\end{equation} 
Comparing the term $2\hbar\Lambda/m\omega_0$ with the value of $\gamma_{\varepsilon}$ given in Eq. \eqref{gamma_lambda}, we notice that $\gamma_{\varepsilon} \ll 2\hbar\Lambda/m\omega_0$ when $\mathsf T\gg \hbar \omega/k_\text{\tiny B} \approx 10^{-6}\,$K. If we work in this limit, we can safely neglect $\gamma_{\varepsilon}$ in Eq.~\eqref{nu0} and we have that $\tilde{\nu}_\text{\tiny min}<1/2$, thus indicating that the system is in an entangled state, when $2\hbar\Lambda/m\omega_0<\eta\omega_0$. Notably, this latter condition can be rewritten as
\begin{equation}
\Lambda<\frac{Gm^2}{\hbar d^3},
\label{env_krisnanda}
\end{equation}
which coincides with the condition in Eq.~\eqref{env_bose} in the limit of $\Delta x\ll d$ once one considers the short-wavelength approximation in Eq.~\eqref{gamma_deco}, i.e.~$\Gamma\simeq \Lambda \Delta x^2$ which is valid for $\Delta x\ll a$. Thus, both Eq.~\eqref{env_bose} and Eq.~\eqref{env_krisnanda} can be synthesised as Eq.~\eqref{tau_bose}: to generate gravitational induced entanglement,  the characteristic time of entanglement creation must be shorter than the decoherence time characterizing the interaction of the system with its surrounding environment.\\

\begin{figure}[t]
	\centering
	\includegraphics[scale=0.35]{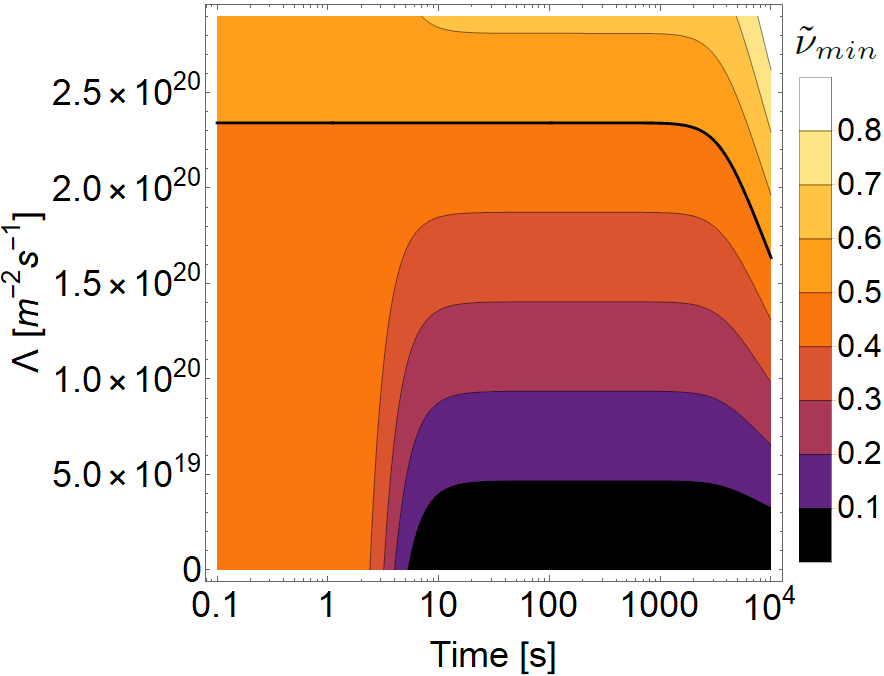}
	\includegraphics[scale=0.7]{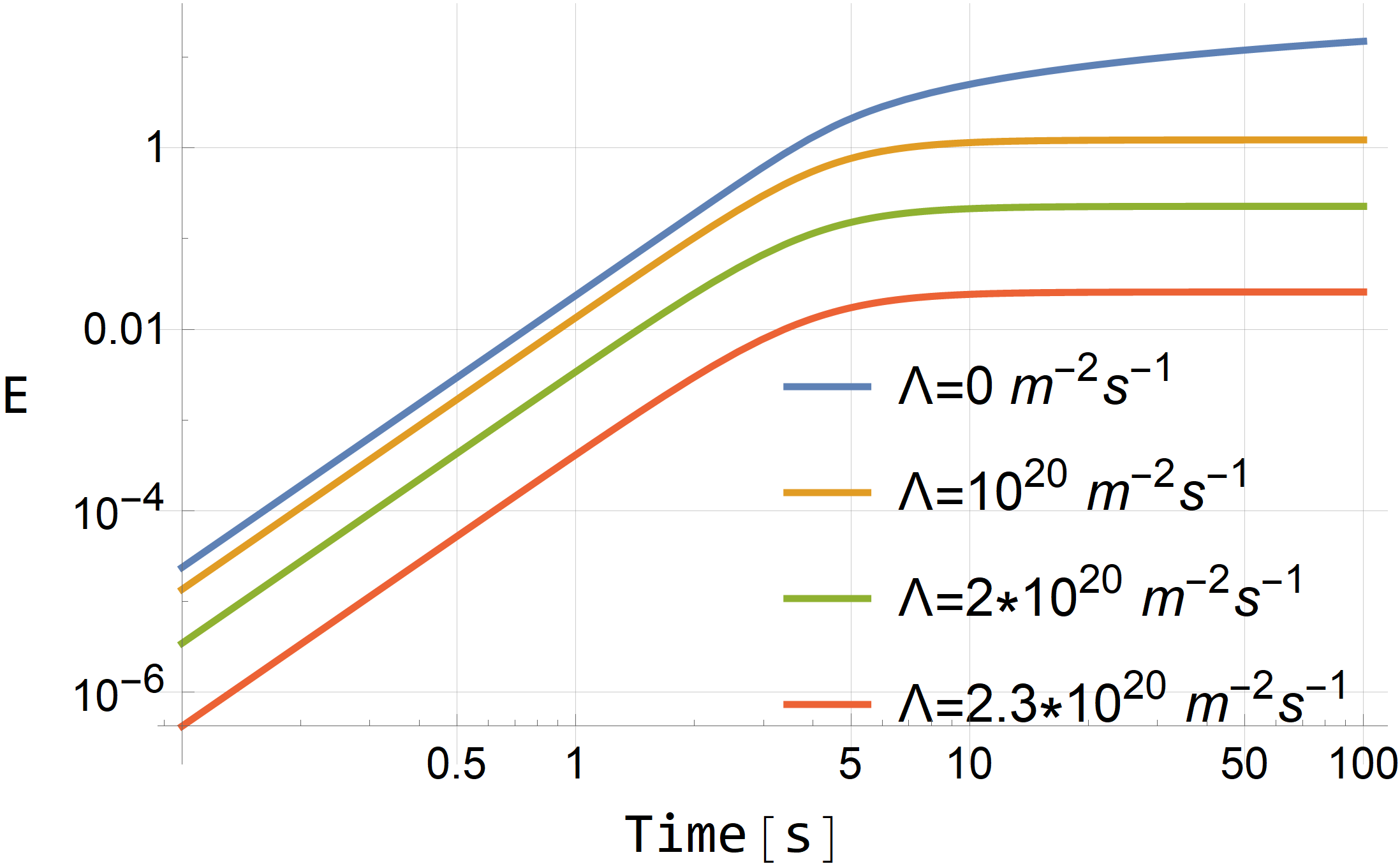}	
	\caption{(Top panel) Minimum symplectic eigenvalue $\tilde{\nu}_{min}$ of the covariance matrix of $\tilde{\rho}$ for different values of time and $\Lambda$, with $\gamma_\varepsilon=0$. The region $\tilde\nu_\text{min}<0.5$, below the black curve, corresponds to entangled states of the system.
	Values lower than 0.5, which is highlighted with a black line, correspond to entangled states of the system. A numerical analysis shows that the results do not change significantly for values of $\gamma_\varepsilon$ up to $10^{-5}\,\text{s}^{-1}$, which would correspond to pressures around $10^{-2}\,$Pa and temperatures around $500\,$K.
	(Bottom panel) Evolution of the logarithmic negativity  $E$ as reported in Eq.~\eqref{logneg} for different values of $\Lambda$. As for the top panel, we consider $\gamma_\varepsilon=0$. }
	\label{krisnanda_env}
\end{figure}

\begin{figure}[t]
	\centering
	\includegraphics[scale=0.75]{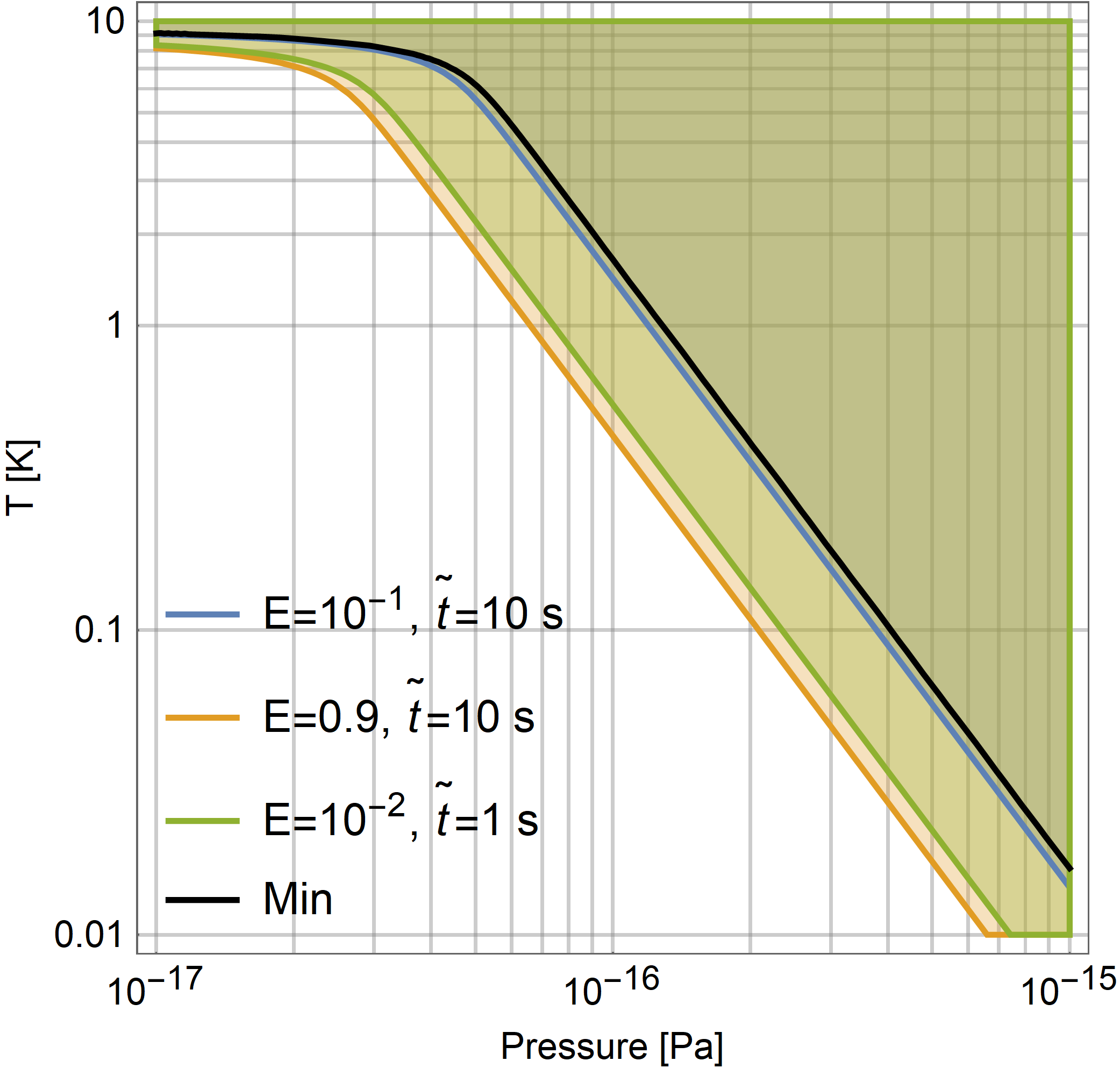}
	\caption{Environmental conditions required to observe different amounts of entanglement in the setup of Krisnanda \textit{et al}. Each line represents the necessary environmental conditions required to observe a given value of logarithmic negativity $E$, as reported in Eq.~\eqref{logneg}, at a fixed time $\tilde{t}$. The shaded areas correspond to values of pressure and temperature that prevent the observation of the targeted value of $E$ within  time $\tilde{t}$. \textit{Min} denotes the minimum experimental conditions required to have any amount of entanglement in the system. The residual gas is assumed to be helium.}
	\label{pT_krisnanda}
\end{figure}

For later times, the behaviour of the entanglement can be seen in the upper panel of Fig.~\ref{krisnanda_env} where we report $\tilde{\nu}_\text{\tiny min}$ for different values of time and of $\Lambda$, with $\gamma_\varepsilon=0 $. A numerical analysis shows that the results do not change appreciably for values of $\gamma_\varepsilon$ up to $10^{-5}\,\text{s}^{-1}$, which correspond to pressures around $10^{-2}\,$Pa and temperatures around $500\,$K. In the bottom panel of Fig.~\ref{krisnanda_env} we show the time evolution of the logarithmic negativity $E_\text{\tiny K-dec}$ of the system for different values of $\Lambda$.
Finally, in Fig.~\ref{pT_krisnanda} we show the environmental conditions required to obtain a specific amount of entanglement at fixed times with the parameters of Tab.~\ref{par}.  We notice that one needs temperatures lower than $9\,$K and residual air pressures around $10^{-16}\,$Pa to have entanglement. With these pressures and temperatures, considerable amounts of entanglement are reached in about $10\,$s.

\section{CSL MODEL}\label{4}
In addition to environmental decoherence, we study also the possible effects of the Continuous Spontaneous Localization (CSL) model on the two setups. This is the most studied among collapse models  \cite{bassi2003,bassi2013}. The CSL model proposes a non-linear and stochastic modification of the Schr\"odinger equation. This modification induces the collapse of the wavefunction for macroscopic systems 
  while it leaves the dynamics of microscopic systems almost unaffected \cite{bassi2003,bassi2013}. The collapse of the wavefunction causes the suppression of  
quantum superpositions and thus of any non-classical correlation , such as  entanglement.
 CSL  is quantified in terms of
  two parameters $\lambda$ and $r_\text{\tiny C}$, which describe respectively the frequency of the collapse and its spatial resolution.
Theoretical proposals include
 \cite{ghirardi1986,ghirardi1990,ghirardi1995,adler2007}:
\begin{equation}
r_\text{\tiny C}= 10^{-7}\,\text{m},\quad\text{and}\quad
\lambda=(10^{-17}\div10^{-9})\,\text{s}^{-1}.
\label{est_CSL}
\end{equation}
Larger values of $\lambda$ are experimentally  excluded. For further details  we refer to \cite{vinante2020} and references therein. We now apply CSL to the two setups we are considering.

\subsection{BM proposal}
Since, in the BM proposal, the distance $d$ between the two masses is greater than $r_\text{\tiny C}$ [cf.~Tab.~\ref{par}], one can safely neglect possible correlations of the noise and describe the CSL effect by simply using Eq.~\eqref{master_deco}. According to CSL, $a=\rC$, while $\Gamma_0$ depends on the geometry of the system. In particular, for a homogeneous sphere of mass $m$ and radius $R$, $\Gamma_0$ reads \cite{vinante2016}:
\begin{equation}
\Gamma_0^\text{\tiny CSL}=\lambda \frac{6 m^2   r_\text{\tiny C}^4 }{m_0^2 R^4} \left[1-\frac{2r_\text{\tiny C}^2}{R^2}+e^{-\frac{R^2}{r_\text{\tiny C}^2}}\left(1+\frac{2r_\text{\tiny C}^2}{R^2}\right)\right].
\label{gammacsl}
\end{equation} 	
Consequently, by exploiting the relation in Eq.~\eqref{env_bose}, one finds that gravitationally induced entanglement can be generated only for $\lambda\leq  10^{-24}\,\text{s}^{-1}$ at $r_\text{\tiny C}=10^{-7}\,$m.
This value is 7 orders of magnitude smaller than the theoretical estimates on the lower bound on $\lambda$. Therefore we conclude that any value of 
the CSL parameters proposed in the literature would prevent the creation of entanglement in the BM setup. 

\subsection{Krisnanda \textit{et al.} proposal}
For the setup of Krisnanda \textit{et al.}, we quantify the effect of the CSL model using the Langevin equations \eqref{lang}, where we substitute the second equation with \cite{bahrami2014,vinante2016}:
\begin{equation}
\frac{\D\hat{p}_j(t)}{\D t} =  \frac{i}{\hbar}\left[\hat{p}_j(t),\hat{H}\right]  +{\xi^\text{\tiny CSL}(t)}  , \label{lang_CSL} 
\end{equation}
where $\xi^\text{\tiny CSL}(t)$ is a real-valued white noise with zero mean and correlation function reading \cite{carlesso2016a}
\begin{equation}
\braket{ \xi^\text{\tiny CSL}_i(t)\xi^\text{\tiny CSL}_j(t') }=2\Lambda^\text{\tiny CSL}\hbar^2\delta (t-t')\delta_{ij},
\label{corr_CSL}
\end{equation}
with $\braket{\cdot}$ denoting the average over the realizations of the noise, and $\Lambda^\text{\tiny CSL}= \Gamma_0^\text{\tiny CSL}/4r_\text{\tiny C}^2$. Since the distance between the two spheres is much larger than $r_\text{\tiny C}$, the CSL noise acts independently on the two masses. Accordingly to CSL, there is no dissipation. Therefore, Eq.~\eqref{env_krisnanda}, where $\Lambda$ is substituted by $\Lambda^\text{\tiny CSL}$, holds true for small times and defines the regime where gravitationally induced entanglement can be generated.
By using the values of the parameters in Tab.~\ref{par}, the latter condition is satisfied for $\lambda\leq 10^{-23}\,\text{s}^{-1}$ at $r_\text{\tiny C}=10^{-7}\,$m, which is 6 orders of magnitude smaller than the lower value in Eq.~\eqref{est_CSL}. Therefore, the same conclusion as for the BM proposal holds: the presence of the CSL noise would prevent the creation of entanglement in the setup.

\section{CONCLUSIONS}\label{5}
\begin{table}[b]
\caption{{Free-fall times $t$ and heights $h=\tfrac12gt^2$, with $g\simeq9.8\,$m/s$^2$, required to generate the amount $E$ of entanglement at fixed values of temperature $\mathsf T$ and pressure $P$ for the proposals of BM and Krisnanda. }}
\label{tabcomparison}
	\centering
	\begin{tabular}{c|ccccc}
		\hline
		\hline
		Proposal 	& $\mathsf T\,$[K]&$P$\,[Pa]	&$E$	& $t$\,[s]	&$h$\,[m] \\ 
		\hline
		\multirow{4}{*}{BM}		&$1$&$10^{-16}$	&$10^{-2}$&$0.15$		&$0.1$\\
				&$1$&$10^{-16}$	&$10^{-1}$&$1.5$		&$11$\\
				&$1$&$10^{-15}$	&no generation&\cancel{\phantom{$1$}}		&\cancel{\phantom{$1$}}\\
				&$10^{-2}$&$10^{-15}$&no generation&\cancel{\phantom{$1$}}		&\cancel{\phantom{$1$}}\\
				\hline
			\multirow{4}{*}{Krisnanda}	&$1$&$10^{-16}$	&$10^{-2}$&$1.1$		&$6.2$\\
				&$1$&$10^{-16}$	&$10^{-1}$&$2.9$		&$42$\\
				&$1$&$10^{-15}$	&no generation&\cancel{\phantom{$1$}}		&\cancel{\phantom{$1$}}\\
				&$10^{-2}$&$10^{-15}$&$10^{-2}$&$1.2$		&$7.6$\\
		\hline
		\hline
	\end{tabular}
\end{table}

In this paper, we have studied decoherence effects on the gravitationally induced entanglement in the setups of BM \cite{bose2017,marletto2017c} and Krisnanda \textit{et al.} \cite{krisnanda2020}. We have found an analytic expression for the entanglement when decoherence is explicitly considered in the dynamics of the system.
{Moreover, we have determined the maximum strength of decoherence, quantified by the expression in Eq.~\eqref{tau_bose}, which allows for gravitationally-induced entanglement generation in these setups. Such an expression also provides the ground to set the requirements of temperature and pressure necessary to perform the experiment. 

The numerical analysis has shown that temperatures and pressures as low as $1\,$K and $10^{-16}\,$Pa are sufficient for generating entanglement with $E=10^{-2}$ after $0.15\,$s for the BM proposal and after $1.1\,$s for the Krisnanda proposal. Notably, cryogenic experiments can easily provide  temperatures reaching 10\,mK and
pressures down to $10^{-16}\,$Pa have already been reached in experiments with Penning traps \cite{sellner2017}.
 However, the time-scale involved requires a free-falling particle to fall for $10\,$cm and $6.2\,$m respectively for the BM and the Krisnanda proposals. Clearly, maintaining the above conditions of temperature and pressure can be technically challenging. As pointed out in \cite{bose2017}, milder environmental conditions such as $P=10^{-15}\,$Pa would lead to a decoherence time of the order of a few seconds, which is comparable to the time of the experiment. However, such conditions would invalidate the  BM proposal, while the Krisnanda proposal would work only for lower temperatures $\leq10$\,mK.
 Moreover the two particles should remain aligned during the experiments, one should also prepare the system without any horizontal and vertical relative velocity, which, over time, would change the relative distance thus potentially disrupting the experiments. Table \ref{tabcomparison} compares the free-fall times and corresponding heights necessary for the two proposals at different environmental conditions.}
While entanglement could be difficult to generate, one could rely on other non-classical correlations such as discord \cite{carlesso2019}, but this was beyond the aim of the current work.

In addition to environmental decoherence, we have also considered the effect of the CSL model on the two setups. We have shown that any proposed value of the CSL parameters would prevent the creation of entanglement in the two setups. We note that this was previously noted by Bose \textit{et al.} \cite{bose2017} for the BM proposal. Conversely, if gravitationally induced entanglement were observed in such setups, it would rule out the CSL model by $6\sim7$ orders of magnitude. Such an achievement would improve the current bounds \cite{vinante2020} by more than 13 orders of magnitude, showcasing  the necessity of a strong improvement of the state-of-art technology to successfully perform these experiments. On the contrary, one expects no effects from a stochastic gravitational wave background \cite{toros2020b}. 

Finally, given the generality of the approaches used to consider explicitly the decoherence effects in the dynamics of the system, the results that we have obtained can be easily generalized to similar setups \cite{christodoulou2018,christodoulou2020} or to different decohering mechanisms, such as gravitational decoherence \cite{bassi2017}.

\section*{Acknowledgements}
SR is grateful to Tomasz Paterek for useful discussions.
MC and AB acknowledge financial support from the EU H2020 FET project TEQ (Grant No. 766900). AB acknowledges financial support from the COST Action QTSpace (CA15220), INFN and the Foundational Questions Institute (FQXi). This publication was made possible also through the support of the ID 61466 grant from the John Templeton Foundation, as part of the The Quantum Information Structure of Spacetime (QISS) Project (qiss.fr). The opinions expressed in this publication are those of the authors and do not necessarily reflect the views of the John Templeton Foundation.

\bibliography{decoherence_effects}{}


\appendix

\section{Entanglement for the BM proposal} \label{appA}

Here, we derive the eigenvalues of the partially trasposed density matrix in the case of the BM proposal. The analysis is performed  in the case of absence and presence of decoherence effects.

\paragraph{Case with no decoherence.}
If we assume that the initial state of the system is $ (\ket{\uparrow}_1+\ket{\downarrow}_1)\otimes(\ket{\uparrow}_2+\ket{\downarrow}_2)/2$, the initial density matrix in the  $\{\ket{i}_1\otimes\ket{j}_2,\, i,j=\uparrow,\downarrow\}$ basis reads:
\begin{equation}
	\rho(0)=\frac{1}{4}\begin{pmatrix}
		1 & 1 & 1 & 1 \\
		1 & 1 & 1 & 1 \\
		1 & 1 & 1 & 1 \\
		1 & 1 & 1 & 1 \\
	\end{pmatrix}.
\end{equation}
The density matrix at time $t$ is given by $\rho(t)=e^{-iHt/\hbar}\rho(0)e^{iHt/\hbar}\rho$ which, using the  the Hamiltonian of Eq. \eqref{H_bose}, reads:
\begin{equation}
	\rho(t)=\frac{1}{4}\begin{pmatrix}
		1 & e^{-i\Delta_{0-}t} & e^{-i\Delta_{0+}t} & 1 \\
		e^{i\Delta_{0-}t} & 1 & e^{-i\Delta_{-+}t} & e^{i\Delta_{0-}t} \\
		e^{i\Delta_{0+}t} & e^{i\Delta_{-+}t} & 1 & e^{i\Delta_{0+}t} \\
		1 & e^{-i\Delta_{0-}t} & e^{-i\Delta_{0+}t} & 1 \\
	\end{pmatrix} ,
\end{equation}
Let us consider the partial transposition of $\rho$ with respect to the second system:
\begin{equation}
	\tilde{\rho}(t)=\frac{1}{4}\begin{pmatrix}
		1 & e^{i\Delta_{0-}t} & e^{-i\Delta_{0+}t} & e^{-i\Delta_{-+}t} \\
		e^{-i\Delta_{0-}t} & 1 & 1 & e^{i\Delta_{0-}t} \\
		e^{i\Delta_{0+}t} & 1 & 1 & e^{-i\Delta_{0+}t} \\
		e^{i\Delta_{-+}t} & e^{-i\Delta_{0-}t} & e^{i\Delta_{0+}t} & 1 \\
	\end{pmatrix} ,
\end{equation}
The eigenvalues of $\tilde{\rho}$ can be easily evaluated and are those reported in Eq. \eqref{eigen_bose_free}.

\paragraph{Case with decoherence.}
When the system is subject to decoherence, the density matrix of the system at time $t$ can be calculated using Eq.~\eqref{master_deco}:
\begin{equation}
	\rho(t)=\frac{1}{4}\left(\begin{smallmatrix}
		1 & e^{-i\Delta_{0-}t-\Gamma t} & e^{-i\Delta_{0+}t- \Gamma t} & e^{-2\Gamma t} \\
		e^{i\Delta_{0-}t-\Gamma t} & 1 & e^{-i\Delta_{-+}t-2\Gamma t} & e^{i\Delta_{0-}t-\Gamma t} \\
		e^{i\Delta_{0+}t-\Gamma t} & e^{i\Delta_{-+}t-2\Gamma t} & 1 & e^{i\Delta_{0+}t-\Gamma t} \\
		e^{-2\Gamma t} & e^{-i\Delta_{0-}t-\Gamma t} & e^{-i\Delta_{0+}t-\Gamma t} & 1 \\
	\end{smallmatrix} \right).
\end{equation}
Again, if we consider the partial transposition of $\rho$ with respect to the second system we get:
\begin{equation}
	\tilde{\rho}(t)=\frac{1}{4}\left(\begin{smallmatrix}
		1 & e^{i\Delta_{0-}t-\Gamma t} & e^{-i\Delta_{0+}t- \Gamma t} & e^{-i\Delta_{-+}t-2\Gamma t} \\
		e^{-i\Delta_{0-}t-\Gamma t} & 1 & e^{-2\Gamma t} & e^{i\Delta_{0-}t-\Gamma t} \\
		e^{i\Delta_{0+}t-\Gamma t} & e^{-2\Gamma t} & 1 & e^{-i\Delta_{0+}t-\Gamma t} \\
		e^{i\Delta_{-+}t-2\Gamma t} & e^{-i\Delta_{0-}t-\Gamma t} & e^{i\Delta_{0+}t-\Gamma t} & 1 \\
	\end{smallmatrix} \right),
\end{equation}
whose correspondent eigenvalues are given by Eq. \eqref{eigen_bose_deco}.

\section{Symplectic eigenvalues - Krisnanda \textit{et al.}}\label{appB}

In order to calculate the logarithmic negativity of the system, we first find its covariance matrix. This can be done in the following way.
First, if we define
\begin{align}
	\hat{u}(t) & = \begin{pmatrix}
		\hat{X}_A(t) \\
		\hat{P}_A(t) \\
		\hat{X}_B(t) \\
		\hat{P}_B(t)
	\end{pmatrix},
\end{align}
in terms of which we can rewrite Eqs. \eqref{lang_env} in a more compact way:
\begin{equation}
	\frac{d\hat{u}(t)}{dt}=K\hat{u}(t)+\hat{\ell}(t)+\kappa\,,
	\label{lang_comp}
\end{equation} 
where we introduced
\begin{align}
	K = \begin{pmatrix}
		0& \omega & 0 & 0 \\
		\omega \eta & -\gamma_\varepsilon & -\omega \eta & 0 \\
		0 & 0 & 0 & \omega \\
		-\omega \eta & 0 & \omega \eta & -\gamma_\varepsilon
	\end{pmatrix} ,\,
\hat{\ell}(t)= \begin{pmatrix}
	0 \\
	\hat{\xi}^{\prime}_A \\
	0\\
	\hat{\xi}^{\prime}_B
\end{pmatrix},\,
\kappa= \begin{pmatrix}
	0 \\
	k \\
	0\\
	-k
\end{pmatrix}.
\end{align}
The solution of Eq. \eqref{lang_comp} reads:
\begin{equation}
	\hat{u}(t)=W_+(t)\hat{u}(0)+W_+(t)\int_0^t\,dt'W_-(t')[\hat{\ell}(t')+\kappa]\,,
	\label{sol_lang}
\end{equation}
where $W_{\pm}=e^{\pm K t}$.
This expression can be used to calculate the covariance matrix of the system. Its elements are defined as:
\begin{equation}
\bm{\sigma}_{ij}(t)=\frac{\braket{\hat{u}_i(t)\hat{u}_j(t)+\hat{u}_j(t)\hat{u}_i(t)}}{2}-\braket{\hat{u}_i(t)}\braket{\hat{u}_j(t)},
\end{equation}
where $\braket{\cdot}$ denotes the average over the initial state and over the environmental degrees of freedom if the case with decoherence is considered.
Using Eq. \eqref{sol_lang}, we can calculate $\bm{\sigma}(t)$ as:
\begin{align}
	\bm{\sigma}(t)=&W_+(t)\bm{\sigma}(0)W_+^\intercal(t)+ \nonumber \\
	&W_+(t)\left[\int_{0}^{t}dt'\,W_-(t')DW_-^\intercal(t')\right]W_+^\intercal(t)\,,
	\label{cov}
\end{align}
where $W_{\pm}^\intercal$ denotes the transposed of $W_{\pm}$ and $D= \, \text{diag}(0,\mu \omega_0 ,0,\mu \omega_0)$, with $\mu= 2\hbar\Lambda/m\omega_0^2$.
Now, if we divide $\bm{\sigma}(t)$ in the following $2\times 2$ submatrices:
\begin{equation}
	\bm{\sigma}=\begin{pmatrix}
		\bm{\alpha} & \bm{\gamma}\\
		\bm{\gamma}^{T} & \bm{\beta}
	\end{pmatrix},
	\label{covgen}
\end{equation}
we can calculate the symplectic eigenvalues of the covariance matrix of the partially transposed state in the following way \cite{serafini2017}:
\begin{equation}
	\tilde{\nu}_{\mp}=\frac{\sqrt{\tilde{\Delta} (\bm{\sigma}) \mp \sqrt{\tilde{\Delta} (\bm{\sigma})^2-4\text{Det}\bm{\sigma}}}}{\sqrt{2}}\, ,
	\label{eigen_sigma}
\end{equation}
where we defined
\begin{equation}
	\tilde{\Delta} (\bm{\sigma})=\text{Det}\bm{\alpha}+\text{Det}\bm{\beta}-2\text{Det}\bm{\gamma}.
\end{equation}
The smallest of these symplectic eigenvalues is related to the logarithmic negativity of the system via $E=-\log_2(2\tilde{\nu}_{min})$.
\smallskip

If we assume to start in the ground state of the two harmonic traps, the initial covariance matrix appearing in Eq. \eqref{cov} reads:
\begin{equation}
	\bm{\sigma}(0)=\frac{1}{2}\begin{pmatrix}
		1 & 0 & 0 & 0 \\
		0 & 1 & 0 & 0 \\
		0 & 0 & 1 & 0 \\
		0 & 0 & 0 & 1 \\
	\end{pmatrix}.
\end{equation}
We use $\bm{\sigma}(0)$ to find the solution of Eq. \eqref{cov} and the symplectic eigenvalues of the partially transposed covariance matrix via Eq. \eqref{eigen_sigma}. The general expression for $\tilde{\nu}_{\min}$ is cumbersome but expanding $\tilde{\nu}_{min}$ around $\tau=0$, where $\tau = \omega_0 t$, we get:
\begin{equation}
 	\tilde{\nu}_{min}(\tau)=\frac{1}{2}+ \frac{1}{2}(\mu-\eta-\frac{\gamma_\varepsilon}{\omega_0})\tau + O(\tau^2)\,.
 	\label{nu_tau_gamma}
\end{equation}
Replacing $\tau$ with $\omega_0 t$ in Eq. \eqref{nu_tau_gamma} we get Eq. \eqref{nu0}.

If we neglect $\gamma_\varepsilon$, as discussed in the Sec. \ref{3}, the analytic expression for $\tilde{\nu}_{\min}$ becomes much simpler:
\begin{equation}
	\tilde{\nu}_{min}(\tau)=\frac{1}{8 \sqrt{3}\eta} \sqrt{a(\tau)-\sqrt{a^2(\tau)-b(\tau)}}\,,
\end{equation}
where $a$ and $b$ are given by:
\begin{widetext}
	\begin{multline}
		a(\tau)=2\eta\,\left[4 \eta^2 \left(2 \mu  \tau^3+3 \tau^2+3\right)+2 \eta \left(2 \mu  \tau^3+3 \tau^2+6 \mu  \tau+6\right)-6 \mu ^2 \tau^2+3\right]\cosh \left(2 \sqrt{2\eta} \tau\right)    \\
		-2\eta\,\left[ 4 \eta^2 \left(2 \mu  \tau^3+3 \tau^2+3\right)-2 \eta \left(4 \mu ^2 \tau^4+8 \mu  \tau^3+3 \tau^2+12 \mu  \tau+6\right)+6 \mu ^2 \tau^2+6 \mu  \tau+3\right] \\
		-\sqrt{2\eta}\,\left[ 24 \eta^2 \tau (\mu  \tau+1)+\eta \left(-6 \mu -4 \mu ^2 \tau^3+6 \mu  \tau^2+12 \tau\right)-3 \mu  (2 \mu  \tau+1)\right]\sinh\left(2 \sqrt{2\eta} \tau\right)  ,
	\end{multline}
	\begin{multline}
		b(\tau)= 96\eta^2 \left(\mu ^2 \tau^4+2 \mu  \tau^3+6 \mu  \tau+3\right) \left[8 \mu  \eta^2 \tau+8 \eta^2-\mu ^2-4 \mu ^2 \eta \tau^2+\mu ^2 \cosh \left(2 \sqrt{2\eta} \tau\right)-4 \mu  \eta \tau \right.  \\
		\left.  \hspace{6cm} +\sqrt{2\eta} (2 \eta+1) \mu  \sinh \left(2 \sqrt{2\eta} \tau\right)\right].
	\end{multline}
\end{widetext}
This expression is easier to handle and has been used to plot $\tilde{\nu}_{min}$ and $E$ in Fig. \ref{krisnanda_env}.

\end{document}